\documentclass[epj]{svjour}
\usepackage{graphics}
\begin{document}
\title{Linac-Ring Type Colliders: Second Way to TeV Scale }
\author{Saleh Sultansoy\thanks{{\em e-mail:} saleh@gazi.edu.tr}}
\institute{Department of Physics, Faculty of Arts and Sciences,
Gazi University,
Teknikokullar, 06500, Ankara, Turkey. \\
Institute of Physics, Academy of Sciences, H. Cavid Avenue 33,
370143, Baku, Azerbaijan.}
\date{Received: date / Revised version: date}

\abstract{Main parameters and the physics search potentials of the
linac-ring type lepton-hadron and photon-hadron colliders are
discussed. The THERA (TESLA on HERA), "NLC"-LHC and "CLIC"-VLHC
proposals are considered.
}

\maketitle

\section{Introduction}

An exploration of (multi-)TeV scale at constituent level is the main goal of
High Energy Physics in the foreseen future. At the end of the last centure,
four ways to TeV scale, namely, ring type hadron machines, linear
electron-positron machines, ring type muon colliders and linac-ring type
lepton-hadron colliders were discussed (see [1] and references therein).
Today, we deal with following situation:

\begin{itemize}
\item Hadron colliders. The LHC with 14 TeV center-of-mass energy
will start hopefully in 2007 and a hundred TeV energy VLHC is
under consideration.

\item Linear colliders. The CLIC is the sole machine with more
than 1 TeV energy, and 3 TeV center-of-mass energy is considered
as third stage.

\item Muon colliders. After the boom in 1990's, main activity is
transferred to the $\nu$-factory options.

\item Lepton-hadron colliders. The sole realistic way to
\linebreak (multi-)TeV scale is represented by linac-ring type
machines.
\end{itemize}

Therefore, as the second way to (multi-)TeV scale, linac-ring type
lepton-hadron colliders require more attention of the HEP
community. Referring to reviews [1-4] for more details of these
machines, as well as their additional $\gamma p$, $eA$, $\gamma A$
and FEL $\gamma A$ options, we present here non-conventional
approach to future energy frontiers for HEP. It may be well
possible that, instead of constructing linear $e^+e^-$ colliders
in the first stage, more attention must be paid to realizing
linac-ring type $ep$ colliders with the same electron beam energy
(see Table 1).

\section{Linac-ring type colliders }

Linac-ring type colliders were proposed more than thirty years ago [5].
Starting from the 1980's, this idea has been revisited with the purposes of
achieving high luminosities at particle factories [6-11] and high energies at
lepton-hadron and photon-hadron collisions [1-4]. In the last three years:

\begin{itemize}
\item THERA with $\sqrt{s} = 1\div 1.6$ TeV and $L\sim
10^{31}$cm$^{-2}$s$^{-1}$ had been included in the TESLA TDR [12]
as the most advanced proposal among linac-ring type $ep$ colliders

\item The idea of QCD Explorer (70 GeV "CLIC" on LHC) was proposed
at the informal meeting held at CERN to discuss the possibility to
intersect CLIC with LHC last summer [13]

\item A comparison of $e$-linac and $e$-ring versions of the LHC
and VLHC based $ep$ colliders is performed in [14] and the linac
options are shown to be preferable.
\end{itemize}

\section{Linac-ring $ep$ vs linear $e^+e^-$}

Even a quick glance at Table 1 is enough to be sure that
linac-ring $ep$ colliders may be as important as linear $e^+e^-$
colliders. Although the luminosity of the previous is less than
that of the second by an order, the center-of-mass energy is an
order higher. Considering the Standard Model, if linear lepton
colliders are important for investigation of Higgs mechanism
responsible for electroweak symmetry breaking, then linac-ring
$ep$ colliders have the same importance for investigation of the
region of small $x_g$ at high $Q^2$ which is crucial for QCD. As
for the BSM physics, the linac-ring $ep$ potential is at least
comparable to the potential of corresponding lepton collider.
Although physics potential of the former is not investigated as
well as that of the last, the statement of the previous sentence
can be easily supported by rescaling of the conclusions presented
in [15] where LHC, LEP-LHC and CLIC are compared. All that
mentioned so far, clearly indicates that linac-ring $ep$ colliders
have a unique potential both for the SM and BSM physics research.
Moreover, additional $\gamma p$, $eA$, $\gamma A$ and FEL $\gamma
A$ options enforce this potential.

Earlier, the idea of using high energy photon beams, obtained by
Compton backscattering of laser light off a beam of high energy
electrons, was considered for  $\gamma e$ and $\gamma\gamma$
colliders (see [16] and references therein). Then the same method
was proposed for constructing  $\gamma p$ colliders on the base of
linac-ring type $ep$ machines [17]. Rough estimations of the main
parameters of $ep$ and $\gamma p$ collisions are given in [18].
The dependence of these parameters on the distance $z$ between
conversion region and collision point was analyzed in [19], where
some design problems were considered. It should be mentioned that
$\gamma$ options are unique features of linac-ring type
lepton-hadron colliders.

The luminosity estimations presented in the Table 1 are rather
conservative and can be improved further by applying advanced
methods such as "dynamic focusing" proposed in [20]. We believe
that further developments will follow provided that the subject is
taken seriously by accelerator physics community.

\section{THERA (TESLA-HERA) and QCD Explorer}

Three versions of TESLA-HERA based $ep$ collisions are considered
in the TESLA TDR [12]: $E_e$ = 250 GeV and $E_p$ = 1 TeV with
$L=0.4\times 10^{31}$cm$^{-2}s^{-1}$, $E_e$ = $E_p$ = 500 GeV with
$L=2.5\times 10^{31}$cm$^{-2}s^{-1}$ and $E_e$ = $E_p$ = 800 GeV
with $L=1.6\times 10^{31}$cm$^{-2}s^{-1}$. In order to achieve
sufficiently high luminosity at QCD Explorer (QCD-E with
$\sqrt{s}= 1.4$ TeV) modification of CLIC and/or LHC beams is
needed. For example, super-bunch option of the LHC will give
opportunity to reach $L\sim 10^{31}$cm$^{-2}s^{-1}$ with nominal
CLIC parameters [21].

In principle, THERA and QCD-E will extend the HERA kinematics
region by an order in both $Q^2$ and $x$ and, therefore, the
parton saturation regime can be achieved. The SM physics topics
(structure functions, hadronic final states, high $Q^2$ and small
$x_g$ region etc) which can be investigated at THERA are presented
in [12]. It seems that QCD-E will provide better kinematics for
these topics, however detailed studies are needed. The BSM search
capacity will be defined by future results from the LHC. For
example, if the first family leptoquarks and/or leptogluons have
masses less than 1 TeV, they will be produced copiously. In
general, the physics search program of THERA and QCD-E is a direct
extension of the HERA search program.

The  $\gamma p$ option essentially enlarges the THERA potential
(it seems that this option is not promising for QCD-E due to low
energy of electron beam). This option will give a unique
opportunity to investigate small $x_g$ region due to registration
of charmed and beauty hadrons produced via  $\gamma g \to
Q\bar{Q}$ sub-process. Concerning the BSM physics one can mention
resonant production of the first family excited quarks (if their
masses are less than 1 TeV), associate production of gaugino and
first family squarks (if the sum of their masses is less than 0.5
TeV), resonant production of t-quarks due to anomalous
interactions etc.

The $eA$ and $\gamma A$ options of THERA, as well as $eA$ option
of QCD-E will give a unique opportunity to investigate small $x_g$
region in nuclear medium and allow the exploration of a non-DGLAP
hard dynamics in the kinematics where $\alpha_s$ is small while
the fluctuations of parton densities are large [22].

Colliding of TESLA ("CLIC") FEL beam with nuclei from HERA (LHC)
may give a unique possibility to investigate "old" nuclear
phenomena in rather unusual conditions. The main idea is very
simple [1, 23]: ultra-relativistic ions will see laser photons
with energy $\omega_0$ as a beam of photons with energy
$2\gamma_A\omega_0$, where $\gamma_A$ is the Lorentz factor of the
ion beam. The huge number of expected events and small energy
spread of colliding beams will give opportunity to scan an
interesting region with keV accuracy.

\section{"NLC"-LHC}

The center-of-mass energies which will be achieved at different
options of this machine [24] are an order larger than those at
HERA and $\sim 3$ times larger than the energy region of THERA and
LEP-LHC. Certainly, $L_{ep}\simeq 10^{32}$cm$^{-2}s^{-1}$ is quite
realistic estimation for "TESLA"-LHC (the factor 7 comparing to
THERA is straightforward due to larger value of $\gamma_p$ at
LHC). For "CLIC"-LHC, $L_{ep}\simeq 10^{31}$cm$^{-2}s^{-1}$ can be
achieved with super-bunch structure of LHC and nominal parameters
of 0.5 TeV CLIC (higher luminosity will require a modification of
CLIC parameters, too). The $ep$ option, which will extend both the
$Q^2$-range and $x$-range by more than two orders of magnitude
comparing to those explored by HERA, has a strong potential for
both SM and BSM research. Concerning the $\gamma p$ option, the
advantage in spectrum of back-scattered photons and sufficiently
high luminosity ($L_{\gamma p}> 10^{31}$cm$^{-2}s^{-1}$) will
clearly manifest itself in a search for different phenomena. Rough
estimations [1, 2] show that the total capacity of $ep$ and
$\gamma p$ options for direct BSM physics (SUSY, compositness etc)
research essentially exceeds that of a 0.5 TeV linear collider.

In the case of LHC nucleus beam IBS effects in main ring are not
crucial because of larger value of $\gamma_A$. The main principal
limitation for heavy nuclei coming from beam-beam tune shift may
be weakened using flat beams at collision point. Rough estimations
show that $L_{eA}\cdot A> 10^{31}$cm$^{-2}s^{-1}$ can be achieved
at least for light and medium nuclei. For $\gamma A$ option,
limitation on luminosity due to beam-beam tune shift is removed in
the scheme with deflection of electron beam after conversion [19]
and sufficiently high luminosity can be achieved for heavy nuclei,
too. Certainly, nuclei options of "NLC"-LHC will bring out great
opportunities for QCD and nuclear physics research. For example,
$\gamma A$ option will give an opportunity to investigate
formation of the quark-gluon plasma at very high temperatures but
relatively low nuclear density (according to VMD, proposed machine
will be at the same time $\rho$-nucleus collider).

Due to a larger $\gamma_A$ at LHC the requirement on wavelength of
the FEL photons is weaker than in the case of TESLA-HERA based FEL
$\gamma A$ collider. Therefore, the possibility of constructing a
special FEL for this option may be a matter of interest. In any
case the realization of FEL $\gamma A$ colliders depends on the
interest of "traditional" nuclear physics community.

\section{"CLIC"-VLHC}

There are a number of papers devoted to possible $ep$ colliders
based on VLHC [14, 25, 26]. Two $e$-ring type options are
evaluated: $ep$ collisions in VLHC booster [25, 26] and $ep$
collisions in VLHC main ring [25]. The first option is not a
matter of interest because of LEP-LHC and THERA covering the same
energy region. For the second option, where the construction of
180 GeV $e$-ring in the VLHC tunnel is proposed, there are a
number of objections and the most important one is following:
instead of constructing a multi-hundred $km$ $e$-ring it is more
wise to construct a few $km$ $e$-linac with the same parameters
[14].

Concerning high energy frontiers, even 1 TeV $e$-linac ("TESLA",
"NLC/JLC") will provide  $\sqrt{s}_{ep} = 20$ TeV, whereas 3 (5)
TeV CLIC corresponds to  $\sqrt{s}_{ep} = 34$ $(45)$ TeV. Taking
in mind THERA estimations one can expect $L_{ep}\simeq
10^{33}$cm$^{-2}s^{-1}$ for "TESLA"-VLHC, whereas $L_{ep}\simeq
10^{32}$cm$^{-2}s^{-1}$ is rather conservative estimation for
"CLIC"-VLHC. Let me remind that $\gamma p$ option will provide
almost the same center-of-mass energy and luminosity as $ep$
option. Obviously, Linac-VLHC based $ep$, $\gamma p$, $eA$ and
$\gamma A$ colliders will give opportunity to investigate a lot of
particle and nuclear physics topics in a best manner.

\section{Conclusion}

The importance of linac-ring type ep colliders was emphasized by
Professor B. Wiik at Europhysics HEP Conference in 1993 [27].
Following previous article [28], he argued TESLA type linear
accelerator to be used as linac. The argument is still valid for
LHC-based $ep$ collider. As for VLHC-based $ep$ collider, CLIC
type linear accelerator seems to be advantageous, since the energy
of TESLA of reasonable size is less than 1 TeV for the time being.

At the first glance, our way of arguing and conclusions seem to be a bit
unusual. However, it might happen that LHC results will support this approach.
Therefore, linac-ring type lepton-hadron and photon-hadron colliders must be
taken into account as seriously as linear lepton and photon colliders.

\begin{acknowledgement}
This work is partially supported by the Turkish State Planning Organization
(DPT) under the Grant No 2002K120250.
\end{acknowledgement}

\begin{table}
\caption{Energy Frontiers} \small{\begin{tabular}{|c|c|c|c|}
\hline\hline
  Colliders & Hadron & Lepton & Lepton-Hadron \\ \hline
  1990's & Tevatron & SLC/LEP & HERA \\ \hline
  $\sqrt{s}$, TeV & 2 & 0.1/0.2 & 0.3 \\ \hline
  $L$, $10^{31}$ cm$^{-2}$s$^{-1}$ & 1 & 0.1/1 & 1 \\ \hline
  2010's & LHC & "NLC" & "NLC"-LHC \\ \hline
  $\sqrt{s}$, TeV & 14 & 0.5 & 3.7 \\
  \hline
  $L$, $10^{31}$ cm$^{-2}$s$^{-1}$ & $10^3$ & $10^3$ & $1\div 10$
  \\ \hline
  2020's & VLHC & CLIC & "CLIC"-VLHC \\ \hline
  $\sqrt{s}$, TeV & 200 & 3 & 34 \\ \hline
  $L$, $10^{31}$ cm$^{-2}$s$^{-1}$ & $10^3$ & $10^3$ & $10\div 100$ \\
  \hline\hline
\end{tabular}}
\end{table}

\end{document}